\documentclass[prl,longbibliography,notitlepage,nofootinbib]{revtex4-1}
\pdfoutput=1
\usepackage{hyperref}
\usepackage{subfigure}
\hypersetup{colorlinks=true,allcolors=blue}
\usepackage{hyperref}
\usepackage{graphicx} 
\usepackage[utf8]{inputenc}
\begin{document}
\title{Scientists \textit{\textbf{in silico}}?}
\author{Carl McBride}
\email[]{carl.mcbride@ccia.uned.es}
\email[]{Dr.C.McBride@gmail.com}
\affiliation{
Departamento de Ciencias y T\'{e}cnicas Fisicoqu\'{\i}micas,
        Facultad de Ciencias,
        Universidad Nacional de Educaci\'{o}n a Distancia (UNED),
        28040 Madrid,
        Spain. \\
}
\date{November 2, 2017}
\begin{abstract}
The end (for human scientists) is nigh? The posit of this discourse is that the majority, if not all, scientific research will eventually be undertaken by one, or a number of, weak artificial intelligences.
\end{abstract}
\maketitle
\section{Introduction}
\begin{quotation} {\it ``Hope we're not just the biological boot loader for digital superintelligence. Unfortunately, that is increasingly probable"} Elon Musk, twitter (August 3, 2014) \cite{muskTwitter} \end{quotation}
The last three centuries have borne witness to spectacular progress in mathematics and the natural sciences, with developments such as calculus, classical mechanics, 
thermodynamics, quantum theory and general relativity, to name but a few. 
That said, in a classic paper by Eugene Wigner titled the `Unreasonable Effectiveness of Mathematics in the Natural Sciences' \cite{Wigner}
he observed that it is 
{\it ``\ldots not at all natural that laws of nature exist, much less that man is able to discover them"}.
Towards the end of his paper he proposed the `empirical law of epistemology' to account for our ability to understand, to such a surprising degree, the world around us. 
However, in the very long run it is perhaps inevitable that scientists will eventually become incapable of thinking openly enough to have ideas that have the potential to significantly advance  science. 
In  physics, for example, certain serious impasses seem to have been reached. 
Two of the most well known problems are;
the long standing difficulty in 
melding together the somewhat disparate theories of general relativity and quantum mechanics; the last six decades
have still not produced a satisfactory theory for quantum gravity, with competing theories like string theory or loop quantum gravity struggling to get the job done. 
Another problem that is causing headaches is the nature of so-called `dark matter', non-Baryonic matter that is estimated to constitute a staggering 84\% of the mass of the universe \cite{WMAP_9yr}. 
Theories to explain the observations of an almost solid-body like motion of galaxies
originated with Fritz Zwicky applying the virial theorem in the 1930's \cite{Zwicky_1,Zwicky_2}. 
Over eighty years later there is, as yet, no good theory  to explain what is going on, with hypotheses like `weakly interacting massive particles' (WIMPS), 
`modified Newtonian dynamics (MOND),  
and extensions to the Standard Model seemingly ruled out. 
There are now proposals for a `hidden sector', composed of more complex self-interacting dark matter \cite{CDM}, an obscured section of the universe that we can only detect by way of gravity.

The Church-Turing-Deutsch principle states that: `every finitely realizable physical system can be perfectly simulated by a universal model computing machine operating by finite means' \cite{DeutschPrinciple}.
Michael Nielsen has pointed out that
{\it ``No one has yet managed to deduce this form of Deutsch’s principle from the laws of physics. Part of the reason is that we don’t yet know what the laws of physics are!"} \cite{Nielsen}.
If this is indeed the case, and we understand less about physics than we think we do, then scientists may sooner-or-later end up instead dedicating their time to going down rabbit-holes;
for example perhaps by unintentionally extrapolating beyond the validity of their models, or producing work that, although correct, is essentially  irrelevant.  
Adopting the words of Nobel laureate David Gross, the issue
{\it ``\ldots is not one of ideology but strategy: What is the most useful way of doing science?"} \cite{Wolchover}.

In Part I of this manuscript I shall focus on the practical aspects of the ongoing project of 
a complete knowledge of science: philosophy, our perceptive abilities, human error and biases, 
working practices, and aesthetics.
In Part II I shall mention five examples of the progress that the (relatively) nascent field of artificial intelligence (AI) is making in science.
\section{Part I}
\subsection{Philosophy of scientists}
It was the illustrious  Auguste Comte, founder of positivism and regarded as  one of the first modern philosophers of science that, in 1835,  famously stated:
{\it ``On the subject of stars, all investigations which are not ultimately reducible to simple visual observations are \ldots 
necessarily denied to  us. While we can conceive of the possibility of determining their shapes, their sizes, and their motions, 
we shall never be able by any means to study their chemical composition or their mineralogical structure \ldots 
Our knowledge concerning their gaseous envelopes is necessarily limited to their existence, size \ldots and refractive power, 
we shall not at all be able to determine their chemical composition
or even their density\ldots
I regard any notion concerning the true mean temperature of the various stars as forever denied to us."} \cite{Comte}.
This statement was evidently made without being aware of the future impact of  the experimental work of Joseph von Fraunhofer, published in 1817 \cite{Fraunhofer} 
providing the basis for the eventual science of spectroscopy - the study of light spectra.
In conjunction with the work of Gustav Kirchhoff on atomic absorption lines \cite{Kirchhoff,KirchhoffEnglish},
by the late 1800's it was indeed possible to have a very good idea of the chemical composition of stars.
With the work of Josef Stefan it was also possible to obtain a value for the temperature of stars.
The point is that one of the greatest intellects ever, by way of a reasoned argument, arrived at a conclusion that nature was 
able to throw back at us for being unfounded.
\begin{figure}
  \includegraphics[width=18cm]{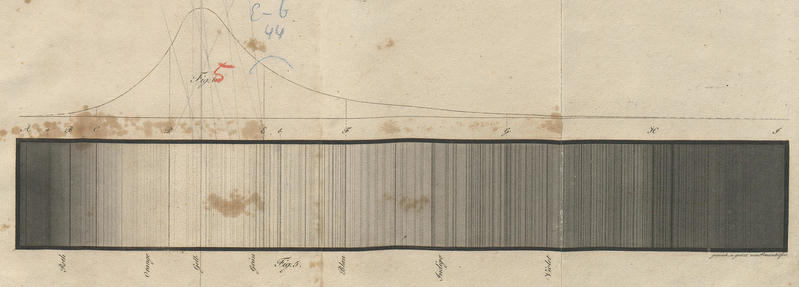}
  \caption{Fraunhofer's adsorption lines in the solar spectrum, taken from his work published in 1817 \cite{Fraunhofer}.}
\end{figure}

Incidents like this have unfortunately lead to 
there being some notable modern-day detractors of the value of philosophy in science; in a recent Google Zeitgeist conference, Stephen Hawking went as far as to say  that ‘philosophy is dead’ \cite{GoogleZeitgeist}.
The physicist Lawrence Krauss said
{\it  ``\ldots the worst part of philosophy is the philosophy of science; the only people, as far as I can tell, that read work by philosophers of science are other philosophers of science. It has no impact on physics what so ever"}
\cite{Krauss}.
Criticisms of the philosophy of science aside, we are interested in epistemology and the acquisition of knowledge.
All scientists, be it consciously or unconsciously, have their own philosophy of science, which
influences their approach to the work that they do.
That said, it is often the case that working scientists  do not have a clear premeditated `philosophy' which guides them in their professional endeavours.
It has been suggested that
most scientists are `critical realists', a subset of scientific realism  \cite{Polkinghorne}.
Scientific realism is the viewpoint that there exists a world independently of ourselves (our minds), and that science
does a progressively good job of describing both the `observable' and the `unobservable' parts of it.
That science as a whole represents an ever improving approximate truth.
An observable would be something one could directly experience, whereas an unobservable would be something
for which there is indirect evidence, such as electrons, quarks etc.
Within scientific realism there are three main variants: explanationist realism; entity realism; and structural realism.
Explanationist realism believes in unobservables if they form part of a theory.
entity realism asks for a stronger justification of unobservables before assigning them to be part of reality, such as being able to causally manipulate them,
and structural realism, which believes in the structure, but does not assign reality to unobservables.
As well as the realists, there are are the anti-realists, with a dual viewpoint which can be divided up into empiricists, constructivists, operationalists or instrumentalists.
In a survey in the magazine Physics World, undertaken by Robert P. Crease \cite{Crease_1}
it became evident that many scientists do not have  a clear-cut philosophical approach to their work, and indeed  a number of them
simultaneously maintained traditionally diametric philosophies \cite{Crease_2}.

In Immanuel Kant's `Critique of Pure Reason' (1787) he states:
{\it ``All our knowledge begins with sense, proceeds thence to understanding, and ends with reason, beyond which nothing higher can be discovered in the human mind\ldots"}\footnote{Meiklejohn translation.}.
The novelist F. Scott Fitzgerald famously wrote 
{\it ``the test of a first-rate intelligence is the ability to hold two opposed ideas in the mind at the same time, and still retain the ability to function"}
\cite{ScottFitzgerald}.
An AI could hold an arbitrary number of opposed ideas in its memory at the same time, and test them one by one.
If an AI can be programmed to reason better than ourselves, I see no reason why an AI could not make a better scientist.
It is interesting to speculate whether an AI will be a naïve realist, having access to a Kantian `noumenal' world, 
and the privilege of a vision of an objective reality, seeing {\it the truth, the whole truth, and nothing but the truth}.
\subsection{Reality is an illusion: Interface theory of perception}
\begin{quotation} {\it `` \ldots humans could infer only as much as their senses allowed, but not experience the actual object itself".} Immanuel Kant \cite{KantDissertatoin} \end{quotation}
\begin{figure}[t!]
  \includegraphics[width=7cm]{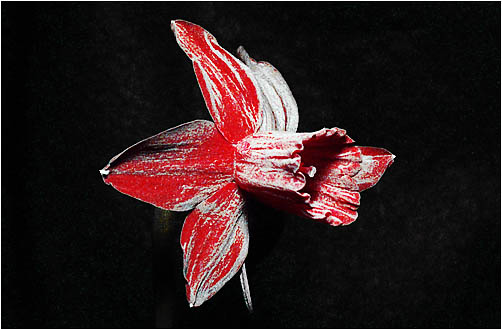}
  \caption{A wild daffodil as seen with an ultraviolet filter (Photo courtesy of Bjørn Rørslett).}
\end{figure}
Let us take for example the humble bee; bees, and many other birds and insects, can see ultraviolet light, and many flowers have a totally different aspect when seen in the ultraviolet (see Fig. 2).
If bees saw as we do, they would almost certainly not survive very long, having great difficulty in locating their food source.
Work by Donald Hoffman and co-workers on the Interface Theory of Perception \cite{ITP}
indicates that we are essentially incapable of correctly viewing reality. They find that
{\it ``veridical perceptions can be driven to extinction by non-veridical strategies that are tuned to utility rather than objective reality"}  \cite{veridical_1}.
They postulate that evolution selects us on the basis of fitness for seeing the world
from the point of view of survival, rather than rewarding `veridical perceptions'.
In other words, our world-view is an internalised description, arrived at due to our animal nature. 
An analogy used is that of the graphical user interface on our computers; we see and interact with icons for files, CDs, printers, programs etc. We do not see 
the true nature of the files. If we could see the `true' nature of the file, say the spiral of tiny distinctly orientated magnetic domains %1's and 0's
imprinted on a hard disk, we would be at a complete loss to get anything done. 
The reality of the string of 1's and 0's is hidden from us, and that is for the better; to be able to decipher 
the raw data would require a lot of time and energy, valuable resources when it comes to survival.
Hoffman creates `fitness functions — mathematical functions that describe how well a given strategy achieves the goals of survival and reproduction'.
These functions are then input into Monte Carlo simulations that run evolutionary `games'.
The results seem to indicate that in the end we only see a `symbolic' version of reality, because that is faster and cheaper and thus favors survival.
\subsection{Human error: logarithms: exponentially difficult?}
\begin{figure}[t!]
  \includegraphics[width=7cm]{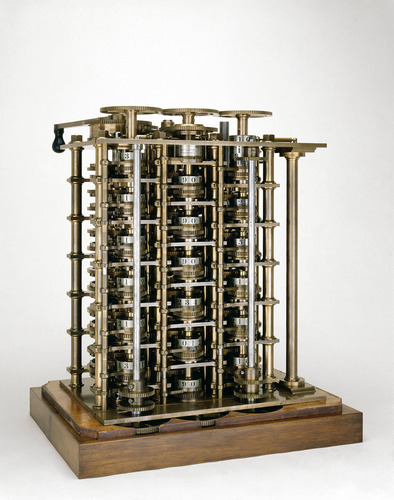}
  \caption{A portion of Babbage's Difference Engine No. 1 (c. 1832) which currently stands in the 
    Science Museum in London. (Source: Science Museum Group Collection Online).}
\end{figure}
The difficulty associated with scientists doing  science could be seen as being somewhat analogous to that of John Napier (as well as others to come). His tables of logarithms
`Mirifici Logarithmorum Canonis Descriptio' (1614) and `Logarithmorum Chilias Prima' (1617) with Henry Briggs,
were calculated by hand over a period of years.
These tables were inevitably error-prone (for example see this analysis \cite{logErrors}),
they had to be; the human mind is not conditioned to calculate error free
to that extent. Such errors were not only limited to the calculations, but also introduced in the transcription and printing process.

A little over two-hundred years later Charles Babbage saw this weakness and invented his difference engine, a mechanical entity whose `world view' is purely numerical.
In 1822 he presented a short paper to the Astronomical Society of London `On the application of machinery to the computation of astronomical and mathematical tables',
and in 1831 he produced the book `Table of the Logarithms of the Natural Numbers from 1 to 108000' \cite{Babbage}.
Unfortunately only small sections of his Difference Engines were actually constructed,
and his later analytical engine was never built, much to the chagrin of his sponsors, the British government.

Today, to attempt to calculate logarithms by hand would be considered a fool's errand.
It is entirely possible that the current situation in the natural sciences is analogous; doing science `by-hand'
is inefficient and error-prone. What is needed this time is some sort of {\it intelligence engine}.
\subsection{Experimenter bias}
Psychologists have long been aware that we play host to a large array of cognitive biases \cite{NotSoSmart},
a number of issues have long been known, in particular a group of phenomena known as `experimenter bias' --although it almost goes without saying that theoreticians can also partake. 
Due to this it can be sometimes exceedingly difficult to set-up a well designed experiment.
In 1979 David Sackett compiled a `catalogue of biases which may distort the design, execution, analysis and interpretation of research'. 
He identified no less than 35 factors that affect results \cite{Sackett}.
A few examples of biases particularly relevant to scientific investigation are mentioned in an article by Regina Nuzzo \cite{Nuzzo}:
\begin{itemize}
\item Hypothesis myopia - where one finds what one is looking for, and ignores other hypothesis.
\item The Texas sharpshooter - misinterpreting random patterns as being meaningful (the origin of the term is the story of a young boy who never misses; he simply shoots at the side of a barn, 
then {\it a posteriori} draws a target around the hole)
\item Asymmetric attention - paying more attention to unusual results than to `expected' results
\item Just-so storytelling - {\it a posteriori} rationalisation of results
\item Stopping data collection before time due to a false positive in A/B tests.
\end{itemize}
Various research protocols can put in place in order to avoid these problems by using mechanisms such as:
\begin{itemize}
\item Devils Advocacy - actively test alternative hypothesis
\item Pre-commitment - publish intentions before performing data collection
\item Teams of rivals - collaborate with groups that hold strongly different opinions
\item Blind data analysis - undertake analysis on an `unknown' data set, without knowing the results (see also \cite{blind})
\item Having, and sticking to, stopping rules for A/B tests
\end{itemize}
\subsection{Publish or perish}
A less philosophical but more practical problem is summed up in
the phrase `publish or perish', which has been around since 1927 \cite{publishPerish} and refers to in implicit pressure to publish academic studies 
in order to obtain tenure, funding, etc.
Indeed, year on year there has been a significant rise in scientific production. 
The open access PubMed database currently indexes over 25 million journal articles in the fields of biomedical and the life sciences. 
Their annual statistical reports \cite{PubMedStats} indicate that in recent years this figure grows by almost a million per annum, corresponding to the addition 
of over one hundred papers an hour.
It is inevitable that not all of this is necessarily quality work.
It is also impossible, even for a whole team of researchers, to be constantly aware of all the research publication relevant to their work
\footnote{Indeed there is now an AI powered search engine, \href{https://www.semanticscholar.org/}{Semantic Scholar}, designed to address this situation.}.
Correspondingly there is an ever increasing rate of retractions in the published scientific literature \cite{vanNoorden}.
The situation is not just the fault of individual scientists.
Recent work \cite{badScience} has  studied the enduring prevalence of false positives in scientific publications
from the viewpoint of natural selection.
To do this a dynamical population model was created whose protagonists had the `utmost integrity', and never cheated
\footnote{The protagonists in the study; simulated scientists, were all treated ethically.}.
Their model consisted of three assumptions:
\begin{itemize}
\item Each laboratory of researchers were capable of positively identifying a true association
\item Unless otherwise, the better this capability, the greater the rate of false positives
\item If effort were to be expended in identifying false positives then productivity would decrease
\end{itemize}
One can then imagine how, in an environment where productivity means prizes, the eventual 
result is the indirect encouragement of `poor methodological practices'.
In some fields, such as biomedical research, the situation has become so bad that it has been suggested that a good many
of studies, for various reasons, are incorrect \cite{false}.
In a recent statistical analysis of publications in the field of anesthesia \cite{Carlisle} it was found that in a set of 5087 clinical trials published between 2000 and 2015,
serious problems were identified in 4.1\% of these papers, indicative of corrupted, incorrect, or just plain falsified data. 
These practical problems, (usually) unintentionally introduced by the scientists themselves, could be viewed as being, in some way, 
similar to the `transcription and printing' problems encountered with the tables of logarithms mentioned earlier.
\subsection{Mathematics and computer-assisted proofs: anti-aesthetic?}
\begin{quotation} {\it ``A good mathematical proof is like a poem -- this is a telephone directory!"} \end{quotation}
\begin{figure}
 \centering
  \subfigure[]{\includegraphics[height=7cm]{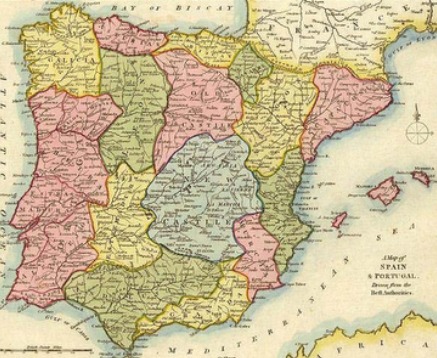}}
   \hspace{1cm}
  \subfigure[]{\includegraphics[height=7cm]{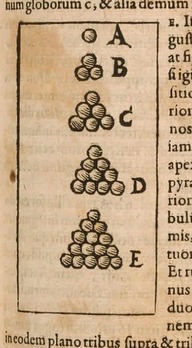}}
  \caption{(a) A map using four colours. 
(b) One of the diagrams in Kepler's `Strena seu de nive sexangula' (Courtesy of The Rare Book \& Manuscript Library of the University of Illinois at Urbana-Champaign).}
\end{figure}
In this section we shall mention two mathematical problems that are easy to formulate, but after many years had passed, were eventually proven in good part with the aid of computers (Note: this is an anathema to many mathematicians).
The above quip was made with regards to the  proof of the four-color map problem, first posed in 1852
`four colours suffice to colour any planar map so that no two adjacent countries are the same colour' \cite{history4color,book4colors}.
This was one of the first ever problems resolved with the help of a computer.
The solution was eventually obtained in 1977, but only after using  over 1,200 hours on an IBM 360  \cite{AppelHaken_1,AppelHaken_2}, the resulting publication was over one-hundred pages long, 
and was accompanied by 465 pages of microfiche \cite{AppelHaken_M1,AppelHaken_M2}.
Appel and Haken, in their popular article in Scientific American \cite{AppelHaken} interestingly observed that
{\it ``In a sense the program was demonstrating superiority not only in the mechanical parts of the task but in the intellectual areas as well".}

Another high profile problem whose proof required the assistance of computers was the Kepler conjecture, which was 
first set down by Kepler himself in 1611 \cite{Kepler1611} in his treatise `Strena seu de nive sexangula' (On the Six-Cornered Snowflake). 
The Kepler conjecture states that  `no packing of congruent
balls in Euclidean three-space has density greater than that of the face-centered cubic packing'.
The proof (by exhaustion) was finally published in 2006, in six sections, occupying a special issue of the journal, and spans over two hundred and sixty pages \cite{HalesFerguson}
This proof was only formally verified in 2014 \cite{Kepler,FlyspeckProject} by what is known as the Flyspeck Project, which made use
of the HOL (Higher Order Logic) Light proof assistant.
However, if there ever was an example on a `non-surveyable proof' it would be the solution of the Boolean Pythagorean triples problem; which takes up a 200 terabyte file \cite{Boolean}. 
A wonderful collection of other examples of computer-assisted mathematics can be found in Ref. \cite{BaileyBorwein}. 

In his 1940 essay `A Mathematician's Apology' G. H. Hardy extols:
\begin{quotation} {\it The mathematician’s patterns, like the painter’s or the poet’s
must be beautiful; the ideas like the colours or the words, must fit
together in a harmonious way. Beauty is the first test: there is no
permanent place in the world for ugly mathematics.}\end{quotation}
This requirement for a proof to be 
{\it ``\ldots elegant, concise and completely comprehensible by a human mathematical mind"} \cite{AppelHaken}, 
or to be {\it like a poem}, a sentiment prevalent amongst mathematicians, has certainly paid off in the past, 
but is not a fundamental requirement, and under some circumstances, could conceivably hinder progress.
\subsection{Our days are numbered}
Our days are numbered, quite literally, in the form of probabilities. Studies have been undertaken to asses the chances that 
various sectors are susceptible to being `computerisable' in the future.
In a study in 2013 \cite{jobs_1} ranking the probability of being computerisable, 
chemists and physicists, both involved in `creative intelligence tasks'
came in at ranks \#173 and \#175 respectively, with a 10\% chance of being `computerisable'
at some point in time, eventually becoming redundant.
Mathematicians fare a bit better, with a ranking of \#135, and a 4.7\% chance of being computerisable.
All other physical scientists come in at \#281, with a 43\% chance.

Indeed, in a 2016 survey, taken amongst 352 of the attendees of two of the most important AI conferences, the consensus was reached that {\it ``there is a 50\% chance of AI outperforming humans in all tasks in 45 years and of automating all human jobs in 120 years"} \cite{GraceEtAl}.
\section{Part II}
\subsection{Artificial Intelligence (AI)}
\begin{quotation} {\it ``Shut up and calculate!"} N. David Mermin \cite{Mermin} \end{quotation}
The above quotation, although taken out of its original context, could well be the apothegm of narrow, or weak, AI. 
The philosopher John R. Searle, in his paper `Minds, brains, and programs',  subdivided artificial intelligence into `strong' (general-purpose) and `weak' \cite{Searle}. 
He describes that in the case of a  strong AI  {\it \ldots ``the appropriately programmed computer really is a mind, in the sense that computers given the right programs can be literally
said to understand and have other cognitive states."}.
On the other hand, a weak AI is a program that has a specific, reduced remit.
For example, a weak AI may be designed for image recognition; it could effectively and efficiently
classify images into categories such as `car', `flower', `cat' etc. A strong AI could do the same thing, 
but at the same time be thinking {\it `I wonder how much that car costs?', `that is a nice flower!', `I am more of a dog than a cat person' \ldots}.

In AI research there is currently no Master Algorithm \cite{Master}, but rather there are five weak `camps':
\begin{itemize}
\item Symbolists, using logic
\item Evolutionaries  with genetic programs \cite{Forrest}
\item Bayesians, with graphical models \cite{Ghahramani}
\item Analogisers, with support vectors
\item Connectionists, with neural networks
\end{itemize}
as yet these threads have not synthesised into one universal approach, which could potentially be `strong'.
\subsection{weak-AI and machine learning}
\begin{figure}
  \includegraphics[width=13cm]{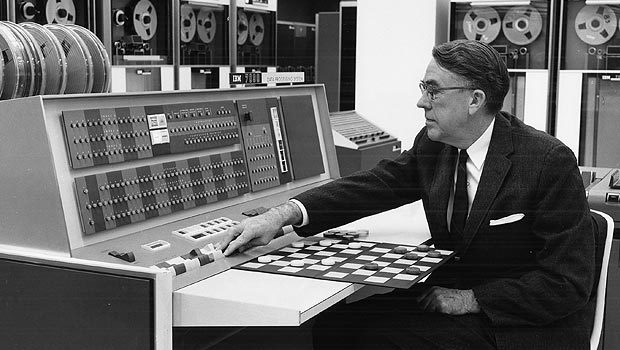}
  \caption{Arthur Samuel and his checkers playing IBM 701 (Photo courtesy of IBM).}
\end{figure}
Machine learning, sometimes known as statistical learning, is a sub-field of AI.
The term `machine learning' was coined by Arthur Samuel in 1959 in his paper describing a computer program designed to play the game of checkers \cite{ArthurSamuel}. 
In his seminal work he combined rote-learning with the groundbreaking `learning-by-generalisation', leading to a program that could play a reasonable game
after only 8-10 hours of running (on an IBM 701).
He defined machine learning as when {\it ``\ldots computers [have] the ability to learn without being explicitly programmed"}.
Computers playing games against humans has long been a testing ground for AI.
For example, a significant milestone was achieved in 1997 when the IBM Deep Blue computer beat the world champion chess player Garry Kasparov \cite{Kasparov}.
In early 2011 IBM performed a live demonstration of a computer system, named Watson, designed to play the TV quiz show game {\it Jeopardy!}.
The novel aspect of {\it Jeopardy!} is that rather than a standard Q\&A format, usually with multiple-choice style answers, 
conversely the  host presents somewhat cryptic answers, and the contestants must formulate the original question that leads to that answer.
Watson played against two of the best previous winners of {\it Jeopardy!} and, in the third and final match,  beat them both.

In 2016 a weak AI hit the headlines again, this time in the form of AlphaGo from Google DeepMind, 
a computer program \cite{AlphaGo} designed to play the ancient Chinese game of Go, a game that was 
previously thought to be, within the foreseeable future, essentially unplayable by a computer.
To the amazement of many, AlphaGo beat 9th dan professional player Lee Sedol by four games to one in a competition consisting of five matches.
More recently Ke Jie, considered the world’s best player, was also beaten by AlphaGo, which won all three games  \cite{TheFutureGoSummit}.
The AlphaGo team have since dispensed with supervised learning, i.e. learning with some form of input from humans, and moved on to pure reinforcement learning.
The latest iteration of the program, AlphaGo Zero \cite{AlphaGoZero} was capable, within three days of being turned on, to become good enough to beat the version that played Lee Sedol
by 100 games to 0. 
Note that despite this amazing feat, AlphaGo Zero is a still weak AI; as it stands it would be completely incapable of playing chess without being reprogrammed by a human.
\subsection{The Technological Singularity}
It was the polish mathematician Stanisław Ulam, in 1958 in his tribute to John von Neumann who mentions \cite{Ulam}
{\it ``the ever accelerating progress of technology and changes in the mode of human life, which gives the appearance of approaching
some essential singularity in the history of the race beyond which human affairs, as we know them, could not continue."}.
In 1966 Irving John Good \cite{Good} described the following scenario:
{\it ``Let an ultraintelligent machine be defined as a machine that can far
surpass all the intellectual activities of any man however clever. Since
the design of machines is one of these intellectual activities, an ultraintelligent
machine could design even better machines; there would
then unquestionably be an “intelligence explosion,” and the intelligence
of man would be left far behind.
Thus the first ultraintelligent machine is the last invention that
man need ever make..."}.,

That said, perhaps the best argument against there ever being  produced such an ultraintelligent machine in the first place was  put forward by Bertram Bowden:
{\it ``there is no point in building a machine with the intelligence of a man, since it is easier to construct human brains by the usual method."}.
\subsection{Five examples of machine learning being used in science today}
\subsubsection{Deep learning: CERN}
One of the `early adopters' of the application of machine learning in scientific discovery was the 
high energy physics community \cite{Denby}, notably at the Conseil Européen pour la Recherche Nucléaire (CERN)
where deep learning \cite{DeepLearning} is being used \cite{Jones} 
to filter through the enormous data sets that are continuously being generated (up to 25 gigabytes per second! \cite{CERN_1}) 
For example, looking for rare exotic particles amongst an enormous sea of particle collisions \cite{Baldi} such as the 
the decay of the Higgs boson \cite{Baldi_2}.
This can be viewed as a classification problem, well suited to deep learning.  
Indeed, {\it ``...deep learning could even lead to the discovery of particles that no theorist has yet predicted"} \cite{CERN_2}.
\subsubsection{Automated experiments: quantum mechanics and MELVIN}
There is perhaps no branch of the physical sciences that is more universally recognised as being counterintuitive than quantum mechanics.
It is a field that pushes our ability to reason, and thus our ability to understand, to the very limits of human intellectual capacity.
In a recent paper \cite{Krenn} a computer program called MELVIN has been used to design configurations that, in the words of the authors 
{\it ``..experiments found by our algorithm show a departure from conventional experiments in quantum mechanics in
that they rely on highly unfamiliar, but perfectly conceivable experimental techniques"}. 
The algorithm starts with a `tool-box' composed of commonly used optical components
readily available in the laboratory such as prisms, mirrors, beam splitters etc. 
It then assembles these components randomly, and learns the output of this hypothetical laboratory setup. 
If the output is deemed to be `useful', it is memorised and can come to form one of the building blocks of a subsequent, more elaborate setup. 
In other words, the algorithm learns from experience.  MELVIN, running for 150 hours (see  \cite{Krenn} for details) 
designed 51 novel, yet feasible experiments. Without learning, the algorithm, running for a period of 250 hours, 
was unable to discover a number of the novelties found with learning.
\subsubsection{Symbolic regression: Eureqa\textsuperscript{\textregistered}}
In 2009 Schmidt and Lipson created a computer program, now known as Eureqa\textsuperscript{\textregistered} \cite{eureqa} 
that was able to re-discover important chunks of classical mechanics by itself.
Directly quoting from the abstract in the Science paper:
{\it ``Without any prior knowledge about physics, kinematics, or geometry, the algorithm discovered Hamiltonians,
Lagrangians, and other laws of geometric and momentum conservation. The discovery rate
accelerated as laws found for simpler systems were used to bootstrap explanations for more
complex systems, gradually uncovering the `alphabet' used to describe those systems"} \cite{SchmidtLipson}.

The Eureqa\textsuperscript{\textregistered} code uses symbolic regression, a technique that is used not only to obtain the parameters 
for an equation from a data set (i.e. traditional regression analysis), but also the equation(s) themselves by way of an evolutionary algorithm  \cite{Forrest}.
The evolutionary algorithm creates trial models by mixing and matching mathematical `genes' (operators, functions etc) to form a `creatures' (equations) well suited to its 
environment (the data set).
\subsubsection{Artificial neural networks: molecular potentials}
In the computer simulation of liquids one has the situation where, lets take as an example  the molecule water, there are at least
one hundred and thirty thermodynamic models currently being studied in the literature \cite{SklogWiki} a good number of which are built upon, or extensions of, 
the parameterised Lennard-Jones model in conjunction with point charges. In the publication \cite{water}
instead neural network potentials, a technique originally designed for brain research, 
were used as a set of `very flexible functions', that were trained 
with the results of a range of condensed phase configurations in order to `learn' the 
{\it ab-initio} potential energy surface of water molecules. 
Such a model is not pre-biased by any prior conceptions of what, in this case a water molecule, should `look-like', no matter how valid the physical reasoning behind the model is.
This is not the only example, artificial neural network potentials have also been successfully developed for 
Al$^{3+}$ ions dissolved in water \cite{GassnerEtAl},
aqueous NaOH solutions \cite{MattiEtAl},
silicon \cite{Cubuk}, gold nanoparticles \cite{ChirikiEtAl},
as well a number of other systems \cite{LorenzEtAl,ManzhosEtAl}.
Jörg Behler has written some very good tutorials as to how to implement molecular potentials in the form of neural networks \cite{Behler1,Behler2}.
\subsubsection{Monte Carlo tree search: organic molecule synthesis}
AI is being applied to the field of organic chemistry. Retrosynthetic analysis involves proposing a 
synthesis route by working backwards to molecules that one knows how to make.
The researchers trained their AI  \cite{AlphaChem} with the Reaxys\textsuperscript{\textregistered} database, which contains over 40 million  chemical reactions,
obtained from patents and publications spanning dating back to 1771.
Using a combination of Monte Carlo tree search in conjunction with a deep neural network, they 
tested their program on 40 randomly selected molecules, finding a restrosynthesys route for 95\% of time, beating
the state of the art Best-First Search, using hand-coded heuristics, which provided routes for only 22.5\% of the molecules in the allotted time. 
\subsection{Possible limitations of AI}
\begin{figure}
  \includegraphics[width=12cm]{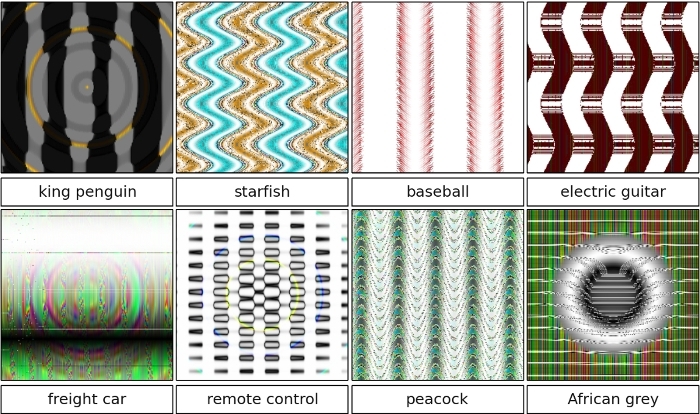}
  \caption{Examples of an AI being fooled into annotating unrecognisable images with extremely high confidence (Source: Ref. \cite{fooled}).}
\end{figure}
It is most likely the case that to do science requires a strong AI, something that some people suggest may be impossible to
create in a computer.
In Searle's aforementioned paper \cite{Searle} he puts forth an argument as to why  strong AI is not possible.
Searl created the Chinese room scenario, which builds upon the famous Turing test \cite{Turing}.
Turing's game essentially involved convincing a human interrogator that he or she is asking questions of a real person and not a computer program, adding that
{\it ``...in order that tones of voice may not help the interrogator the answers should be written, or better still, typewritten. The ideal arrangement is to have a teleprinter communicating between the two rooms."}
In the  Chinese room {\it gedankenexperiment}, Searle supposes that:
\begin{itemize}
\item There is a computer locked in a room that, when fed information written in Chinese, responds satisfactorily via a printout, again in Chinese, in the process easily passing the Turing test 
with its answers. 
\item The computer, being a computer, follows a deterministic algorithm
\item Given time and patience, Dr. Searl could sit in the room instead of the computer, and also follow the algorithm (provided for him in English), and also output satisfactory answers in Chinese
\item Dr. Searl has absolutely no knowledge of Chinese whatsoever, and thus no understanding of either what the input nor the output means
\item Therefore, the computer also has no real understanding of  what it is doing, and as such does not think and has no mind of its own.
\end{itemize}
There have been numerous publications both for and against this argument\footnote{Google Scholar lists over 5,800 citations of the  original paper}.
With regards to `real understanding' Richard Feynman once said about one of the best theories that physicists have developed to date, 
{\it ``... I think I can safely say that nobody understands quantum mechanics."} \cite{Feynman}. 
We may not understand quantum mechanics, but nobody can deny that we certainly spend an inordinate amount of time {\it thinking} about it! 
All said and done, as we have seen with Comte, philosophical arguments can sometimes be overtaken by practical advances.
If the Church-Turing-Deutsch principle \cite{DeutschPrinciple} does hold, then there should be no reason why a strong AI cannot built.
Given that a strong AI in principle can be (and therefore eventually will be) built, here we mention some of
the current problems in AI.

\begin{itemize}
\item The statistical approach of machine learning may not be enough in itself to derive the 
laws of physics and apply them to novel situations. Symbolic learning, using logic, will probably need to form part of 
the AI's make-up in order to have the ability of relational reasoning \cite{relational,relational2}.
\item AI's are not infallible. For example, a task that deep neural networks have become  particularly good at is image recognition.
Each year ImageNet Large-Scale Visual Recognition Challenge is held to test AI's ability to perform various test on a large
database of images, 
and each year the results improve. In 2014, in the image classification part of the challenge the best AI achieved a 6.66\% error rate \cite{ImageNet}, rivalling 
human annotators. 
However, it has been shown that deep neural nets can produce false positives from images (see Fig. 6) that are  meaningless to humans \cite{fooled}.
\item It would be difficult to define a generalised `reward function'. One cannot simply say to an AI `do science',
it needs to know when it is making progress. It some how needs to embody, or more precisely encode, the question that lies  at the heart of science: `Why?'
\item Neural networks, especially deep neural networks,  have been accused of being black-boxes, providing wonderful results, but they themselves are seemingly indecipherable, the so-called `interpretability problem'. 
However, that may be changing, with theories such as the `information bottleneck' method \cite{Tishby,Tishby2}, along with a multitude of mechanisms for teasing out the features that the deep neural network
has honed in  on as being the salient features of, say, an image \cite{Castelvecchi,Voosen}.
\end{itemize}
\section{Epilogue}
Artificial intelligence is ever increasingly finding its way into the scientists tool-box as a powerful technique to aid discovery.
Although there is much ongoing work to develop AIs that perform certain human tasks better than humans can, the biggest rewards will almost certainly come
form AIs performing unthought-of tasks in an unforeseen manner.
However, asking an AI to perform science {\it per se} is a long way off, with 
major advances required before any AI can encapsulate aspects such as motivation and independent creativity, required to tackle such a monumental task. 
That said, eventually we may find ourselves leaving all scientific research to the AI's. 
The predicted `Fourth Industrial Revolution' \cite{4IR} could also bring with it a revolution in the way we undertake science.
When will this `digital superintelligence' that can conduct unsupervised science be developed? According to the Maes-Garreau law
\footnote{Maes-Garreau law: the amusing observation that these type of predictions always seem to coincide with the number of years until the retirement age of the person making it.}, 
sometime in the next 20 years.
Until then we will just have to learn to cooperate with each other \cite{cooperate}.

\begin{quote}
{\it ...``Forty-two!" yelled Loonquawl. ``Is that all you've got to show for
seven and a half million years' work?"\\
--``I checked it very thoroughly," said the computer, ``and that quite
definitely is the answer. I think the problem, to be quite honest with
you, is that you've never actually known what the question is."}\\
 Douglas Adams - The Hitchhiker's Guide to the Galaxy
\end{quote}
\subsubsection{Acknowledgments}
\begin{acknowledgments}
The author would like to thank Cristina Santa Marta Pastrana and Juan J. Freire for their support during the writing of this manuscript.
This work has been supported by a Universidad Nacional de Educaci\'{o}n a Distancia (UNED) Postdoctoral Grant (2013-018-UNED-POST).
\end{acknowledgments}
\bibliography{future}
\end{document}